 \documentclass[final,5p,times,twocolumn]{elsarticle}

\usepackage{graphicx}
\usepackage{amsmath}
\usepackage{amssymb}
\usepackage{caption}
\usepackage{subcaption}
\usepackage{lineno}
\usepackage{xcolor}
\journal{Nuclear Instruments and Methods in Physics Research A}

\begin{document}
	
\begin{frontmatter}
\title{Large Angle MIEZE with Extended Fourier Time}

\author{Ryan Dadisman}
\author{Georg Ehlers}
\author{ Fankang Li \corref{cor1}}
\address{Neutron Technologies Division, Oak Ridge National Laboratory, Oak Ridge, TN 37831, USA}

\fntext[myfootnote]{Note to the editor (not to be printed with the paper should it be published). This manuscript has been authored by UT-Battelle, LLC under Contract No. DE-AC05-00OR22725 with the U.S. Department of Energy. The United States Government retains and the publisher, by accepting the article for publication, acknowledges that the United States Government retains a non-exclusive, paid-up, irrevocable, world-wide license to publish or reproduce the published form of this manuscript, or allow others to do so, for United States Government purposes. The Department of Energy will provide public access to these results of federally sponsored research in accordance with the DOE Public Access Plan (http://energy.gov/downloads/doe-public-access-plan).}

\cortext[cor1]{Corresponding author: frankli@ornl.gov}

\begin{abstract}
{\it M}odulation of {\it I}ntensity {\it E}merging from {\it Z}ero {\it E}ffort (MIEZE) is a neutron resonant spin echo technique which allows one to measure time correlation scattering functions in materials by implementing radio-frequency (RF) intensity modulation at the sample and detector. The technique avoids neutron spin manipulation between the sample and the detector, and thus could find applications in cases where the sample depolarizes the neutron beam. 
However, the finite sample size creates a variance in path length between the locations where scattering and detection happens, which limits the contrast in intensity modulation that one can detect, in particular towards long correlation times or large scattering angles. 
We propose a modification to the MIEZE setup that will enable one to extend those detection limits to longer times and larger angles. 
We use Monte Carlo simulations of a neutron scattering beam line to show that, by tilting the RF flippers in the primary spectrometer with respect to the beam direction, one can shape the wave front of the intensity modulation at the sample to compensate for the path variance from the sample and the detector. 
The simulation results indicate that this change enables one to operate a MIEZE instrument at much increased RF frequencies, thus improving the effective energy resolution of the technique. The simulations show that for an incident beam with maximum divergence of 0.33$^\circ$, the maximum Fourier time can be increased by a factor of 3.
\end{abstract}

\begin{keyword}
MIEZE \sep McStas \sep Phase Correction \sep Tilted RF Flipper
\end{keyword}

\end{frontmatter}

\section{Introduction}
{\it N}eutron {\it S}pin {\it E}cho (NSE) is a method of neutron scattering which encodes the velocity (energy) of the neutron in the Larmor spin precession phase \cite{Mezei1972}
\begin{equation}
\phi=\gamma\int{B}{\;}dt
\label{phi}
\end{equation}
where $\gamma=-1.832\times{10}^{8}{\;}\rm{rad}\cdot{s}^{-1}\cdot{T}^{-1}$ is the gyromagnetic ratio of the neutron and the integral is over the modulus of the magnetic field during the time inside the field. 
A neutron passing through two identical magnetic field regions with $B_1 = - B_2$ will have a total Larmor phase $\phi=0$, so that the final polarization is the same as the initial polarization. If a sample is placed between these two regions and exchanges energy with the neutron, then the neutron velocity and thus time-of-flight through the second magnetic field region will be different, resulting in a non-zero Larmor phase, $\phi\neq{0}$. Using this technique, very high energy resolution can be achieved by measuring the change in the polarization
\begin{equation}
P\propto\cos\left(\frac{\Delta{E}\cdot\tau}{\hbar}\right)
\end{equation}
where $\Delta{E}$ is the energy transfer during scattering, and $\hbar$ is the reduced Planck constant. The Fourier time $\tau$ is a parameter that depends on the field integral, eq.(\ref{phi}), and the third power of the neutron wavelength, $\lambda^{3}$, and is effectively used to tune the energy resolution of the setup. 
A weakness of the NSE method is in the technical difficulties that arise when the sample depolarizes the beam, or when a very large magnetic field is desired at the sample. 
{\it N}eutron {\it R}esonance {\it S}pin {\it E}cho (NRSE) was proposed \cite{Gahler1987} as an alternative to NSE that uses compact RF spin flippers rather than large DC magnetic field regions. 
NRSE enables one to implement the {\it M}odulation of {\it I}ntensity {\it E}merging from {\it Z}ero {\it E}ffort (MIEZE) technique, which performs spin manipulations only before the neutron reaches the sample, and is therefore able to accommodate depolarizing samples and complicated sample environments \cite{GAHLER1992}.
\begin{figure*}[ht]
	\centering
	\includegraphics[width=0.9\linewidth]{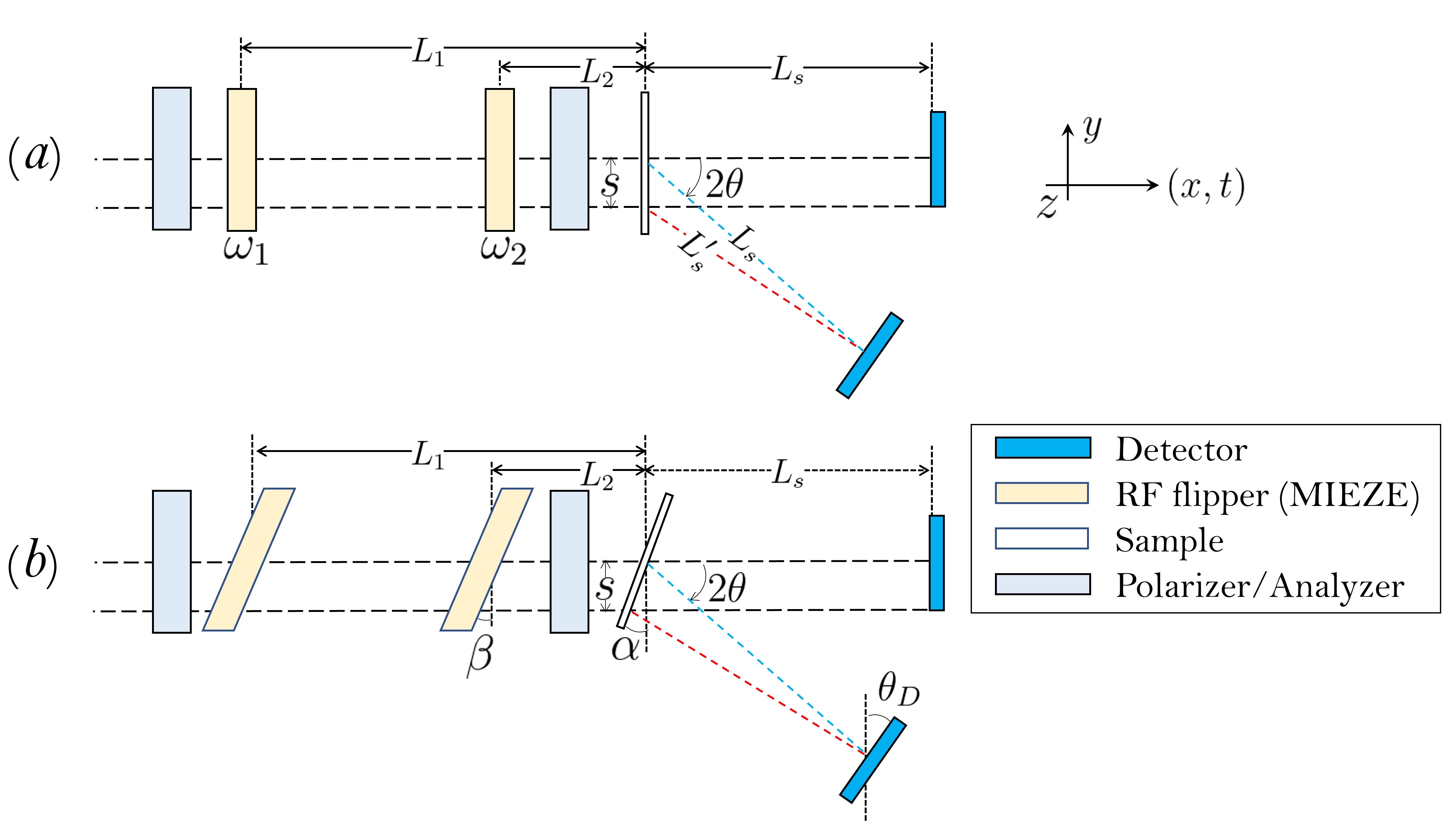}
	\caption{\label{fig:MIEZEsimple} (a) Schematic of a conventional MIEZE setup. (b) Schematic of a MIEZE setup with the RF flippers tilted by an angle of $\beta$ to correct the phase aberration. The sample may be tilted by angle $\alpha$ to achieve a similar effect. $L_1$ and $L_2$ are the distances of the RF flippers to the sample and $L_s$ is the center-to-center distance of the sample to the detector. The two dashed lines between sample and detector denote two scattered neutron trajectories. $\theta_D$ is the tilting angle of the detector from its original direction along $y$.}
\end{figure*}

MIEZE involves a pair of compact RF spin flippers operated at different angular frequencies of $\omega_1$ and $\omega_2$, as shown in Fig.~(\ref{fig:MIEZEsimple}a). The phase of the intensity modulation seen at the detector position is given by,
\begin{eqnarray}
\label{eqn:MIEZEphase}
\phi_{MIEZE} = 2 \Bigg[ (\omega_2 - \omega_1) t  + \frac{ \omega_1 (L_1+L_s)-\omega_2 (L_2+L_s)}{v} \Bigg]{\;},
\end{eqnarray}
where $L_{1,2}$ are the distances from the centers of the two flippers to the sample, $L_{s}$ is the distance from sample to the detector, $v$ is the neutron velocity. As one can see, this phase depends on the neutron time-of-flight ($t$) from the first flipper to the detector. Any variation in the time-of-flight will eventually cause a phase aberration and a loss of contrast in the detector signal.
Such variation could be contributed from the wavelength dispersion, a path length variance, and other effects.  To minimize the phase aberration due to the dispersion in the neutron wavelength (or velocity), the velocity dependent term in eq.~(\ref{eqn:MIEZEphase}) can be minimized by choosing
\begin{equation}
\label{eqn:MIEZEcond}
\frac{\omega_2}{\omega_1} = \frac{L_1+L_s}{L_2+L_s} {\;}{.}
\end{equation}
Eq.~(\ref{eqn:MIEZEcond}) is also called the time focusing condition. With a polarization analyzer after the second RF flipper, which only picks up the polarization vector along the analyzing direction, an intensity modulation in time can be produced, yielding,
\begin{equation}
\label{eq5}
I(t) = A \cos \Big[2 (\omega_2 - \omega_1) t \Big] + C{\;},
\end{equation}
where $A$ is the amplitude of modulations, and $C$ is the time-averaged intensity. Variance in the time-of-flight from the sample to detector due to quasi-elastic scattering will reduce the contrast, which can be used to determine the intermediate scattering function
\begin{equation}
\frac{A}{C} = \frac{S(\mathbf{Q}, \tau_{MIEZE})}{S(\mathbf{Q}, 0)}
\end{equation} 
where $\mathbf{Q}$ is the scattering vector and $\tau_{MIEZE}$ is the Fourier time. For MIEZE, $\tau_{MIEZE}$ is given by
\begin{equation}
\label{eqn:TauMIEZE}
\tau_{MIEZE} = \frac{m^2\lambda^3}{\pi h^2} (\omega_2 - \omega_1) L_s 
\end{equation}
where $m$ is the neutron mass, $h$ is Planck's constant, $L_s$ is the distance from the sample to the detector, and $\lambda$ is the neutron wavelength. 

Another term in eq.~(\ref{eqn:MIEZEphase}) causing a phase aberration is due to the variation of the path length between sample and detector ($L'_s$), as shown in Fig.~(\ref{fig:MIEZEsimple}a).
For the neutron trajectories emerging from the center or the edge of the sample, the difference in the time-of-flight to the detector is given by 
\begin{eqnarray}
\Delta{t} & = & \frac{1}{v}\cdot\left(L_s - L'_s\right)\nonumber\\
 & = & \frac{1}{v}\cdot\left(L_s  - \sqrt{L_s^2 + s^2  - 2s L_s \sin 2\theta}\right)\nonumber\\
 & \sim & \frac{s}{v}\cdot\sin{2}\theta{\;},
\end{eqnarray}
where $s$ is the transverse distance of the scattering event from the centerline, $2\theta$ is the scattering angle, and the last expression is to the lowest order in $s$. 
In practice, this means that MIEZE is limited to studying small samples at low scattering angles, as the product ${s}\cdot\sin{2}\theta$ must be small. 
In the small angle neutron scattering (SANS) regime, one can operate MIEZE at MHz frequency, whereas at larger scattering angles one finds oneself bound by $\omega_{1,2}\sim{2}\pi\cdot{10}$~kHz. 
Previous studies of the resolution function have noted this correlation and concluded that MIEZE is best suited for a SANS configuration~\cite{BRANDL2011,WEBER2013,MARTIN2018}. 

\section{Correcting the Phase Variance of MIEZE}
One may attempt to remove the limitation to stay in the SANS regime by tilting the RF flippers relative to the beam. 
This is a common practice in NRSE work to measure excitation life time of quasi-particles \cite{RN2006}. Alternatively, one may also rotate a flat, disk-like sample relative to the beam. The latter was put forward previously~\cite{BRANDL2011,MARTIN2018}.

\subsection{Tilting the RF flippers}
Consider the two parallel neutron trajectories shown in Fig.~(\ref{fig:MIEZEsimple}b), where one interacts with the center of the sample and the other with the sample's edge at a displacement $s$ from the center. With the RF flippers tilted by an angle $\beta$, and the sample perpendicular to the beam direction ($\alpha=0$), the phase of the neutron spin at the detector is given by 
\begin{eqnarray}
\label{eq9}
\phi_{MIEZE} & = & 2(\omega_2-\omega_1)t\nonumber\\
 & + & \frac{2}{v}\cdot\left[\omega_1(L_1+s\cdot\tan\beta)
 -\omega_2(L_2+s\cdot\tan\beta)\right]\nonumber\\
 & + & \frac{2}{v}\cdot(\omega_1-\omega_2)\cdot{L'_s}\\
 & = & 2(\omega_2-\omega_1)t + \frac{2}{v}\cdot(\omega_1-\omega_2)\cdot{L_s}\nonumber\\
 & + & \frac{2}{v}\cdot(\omega_1-\omega_2)\cdot{s}\cdot\tan\beta\nonumber\\
 & - & \frac{2}{v}\cdot(\omega_1-\omega_2)\cdot{s}\cdot\sin2\theta{\;},
\end{eqnarray}
where ${L_s}-{L'_s}=s\cdot\sin{2}\theta$ from above was inserted.
Thus one can, to the lowest order in $s$, reduce the effect of the sample size by tilting the RF flippers such that
\begin{equation}
\label{eqn:OptRF}
\tan \beta = \sin 2\theta{\;}.
\end{equation}
In combination with the standard MIEZE focusing condition given in eq.~(\ref{eqn:MIEZEcond}), the phase variance due to a finite size sample is minimized (for a parallel beam) at a scattering angle beyond the SANS regime. 

\subsection{Tilting the sample} 
This was studied analytically in refs.~\cite{BRANDL2011} and \cite{MARTIN2018}, and the optimal sample rotation angle was found to be impractical for small angle scattering. Here we study whether this is a practical method at very large scattering angles. Therefore, we will briefly discuss the effects of rotating the sample. Consider the setup shown in Fig. (\ref{fig:MIEZEsimple}b), where the sample is tilted by an angle $\alpha$ and the RF flippers are not tilted.

The phase at the detector for a neutron trajectory with a distance $s$ from the center of the sample is now given by
\begin{eqnarray}
\phi_{MIEZE} & = & 2(\omega_2-\omega_1)t\nonumber\\
 & + & \frac{2}{v}\cdot\left[\omega_1(L_1-s\cdot\tan\alpha)
 -\omega_2(L_2-s\cdot\tan\alpha)\right]\nonumber\\
 & + & \frac{2}{v}\cdot(\omega_1-\omega_2)\cdot{L'_s}{\;},
\end{eqnarray}
which is similar to eq.~(\ref{eq9}). Note, however, that the angle $\alpha$ enters with the opposite sign. 
To the first order, one now has 
\begin{equation}
L_s-L'_s = \frac{s}{\cos\alpha}\cdot\sin\left(2\theta-\alpha\right){\;},
\end{equation}
and from this it follows that the leading order term proportional to $s$ is canceled by satisfying the following condition
\begin{equation}
\label{eqn:OptSampAngle}
-\sin\alpha = \sin\left(2\theta-\alpha\right){\;}.
\end{equation}
This condition illustrates that, when $2\theta$ is outside the SANS regime, the optimal angle $\alpha=\theta\pm\pi/2$ dictates reflection geometry for a disk-like flat sample which becomes `practical' when $2\theta\gtrsim{60}^{\circ}$. 
Otherwise, $\alpha$ is too large and most of the beam passes on the side of a `typical' sample without intercepting it.

\section{McStas Simulation of the Phase Correction}
\subsection{McStas component of the RF neutron resonant spin flipper}
To validate this method, simulations were performed using the neutron beamline simulation software McStas~\cite{McStas99,McStas2020}. 
An RF spin flipper component was coded in-house, which uses a time-dependent magnetic field of the form
\begin{equation}
\vec{B} = 
\begin{pmatrix}
-B_{RF}\sin(\omega t) \\
B_{RF}\cos(\omega t) \\
B_0+B_g\sin(\frac{\pi x}{l})
\end{pmatrix}
\label{eqn:McStasField}
\end{equation}
where $B_0$ and $B_g$ are the static field and gradient field respectively. 
By turning $B_g$ on and off, one can switch between adiabatic ($B_g\neq{0}$) and non-adiabatic ($B_g={0}$) modes. 
The boundaries of the flipper are defined as $x=\pm{l}/2$ along the length $l$ of the flipper. The angular frequency of the flipper is $\omega$ and the amplitude of the RF field is $B_{RF}$. The $\hat{x}$ axis is along the beam direction, and so the amplitudes of the RF and gradient fields always vary as a function of $x$. 
The fields $B_0$ and $B_g$ in eq. (\ref{eqn:McStasField}) are perpendicular to the beam direction which represents the {\it T}ransverse {\it N}eutron {\it R}esonant {\it S}pin {\it E}cho (TNRSE) mode. 
In the code, the vector components of the magnetic field in eq.~(\ref{eqn:McStasField}) can be reconfigured differently so that $B_0$ is parallel to the beam direction ($x$), which corresponds to the {\it L}ongitudinal {\it N}eutron {\it R}esonant {\it S}pin {\it E}cho (LNRSE) mode. 

\begin{figure}[ht]
	\centering
	\includegraphics[width=0.7\linewidth]{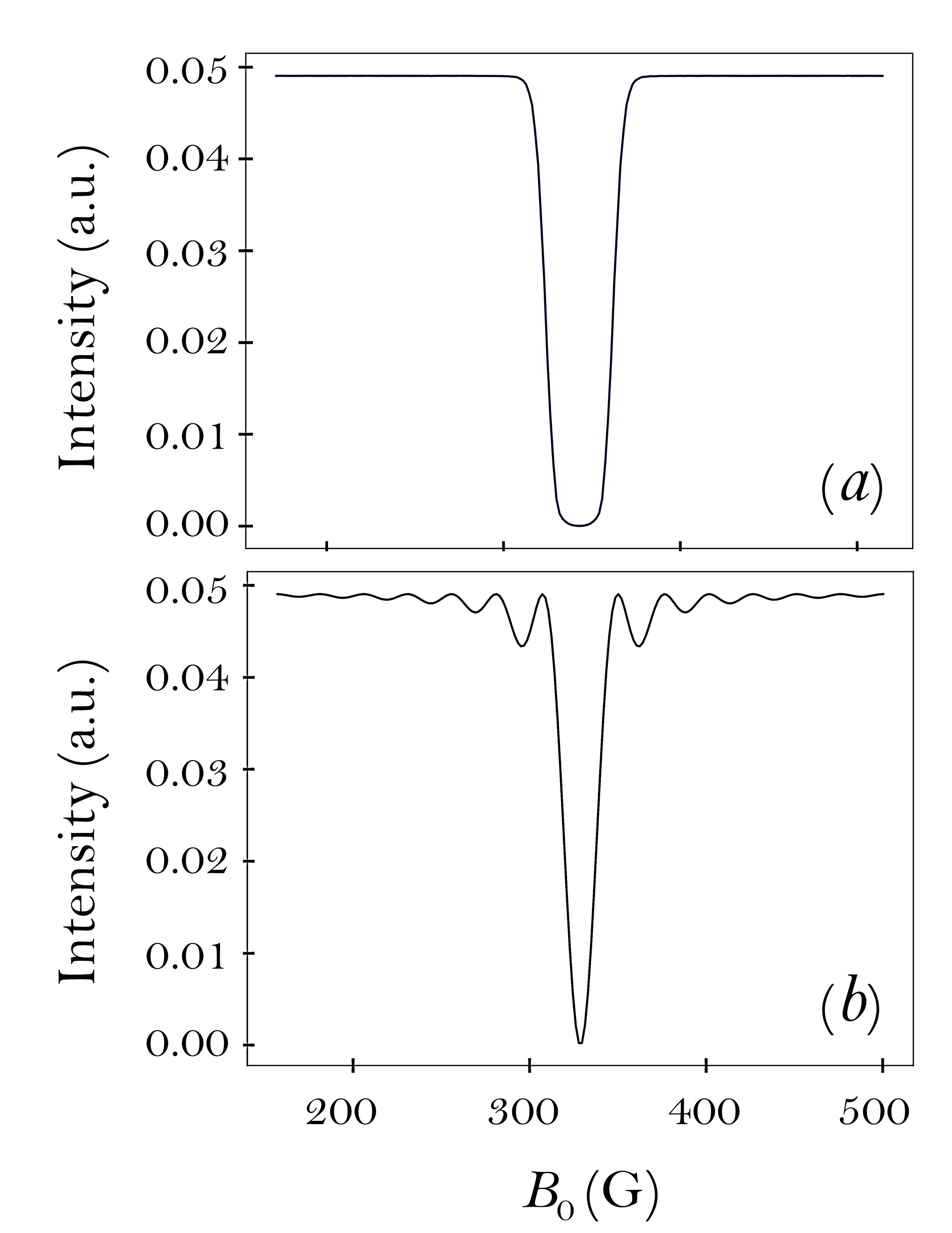}
	\caption{\label{fig:McStasFlipper} Intensity of one spin state by scanning the $B_0$ field, for an RF flipper operating at 1~MHz and 5.5~{\AA} neutrons, where a $B_0={342.8}$~G is expected. (a) Adiabatic mode, with $B_g=B_{RF}=20$~G; (b) Non-adiabatic mode, with $B_g=0$~G and $B_{RF}=1.23$~G. }
\end{figure}

In the following, the transverse configuration is used as in refs.~\cite{LI2020, Dadisman2020}. 
Fig.~\ref{fig:McStasFlipper} shows the simulation results of the spin flipper operating in non-adiabatic and adiabatic spin flipping modes when scanning $B_0$. With the neutron gyromagnetic ratio ($\gamma$) known previously, for a RF frequency of $f$=1~MHz, a resonance peak at $B_0=\frac{2\pi f}{\gamma}=342.8~$G is expected for both modes. The resonance field from the simulation agrees well with the expectation and one can also see that the resonance peak of the adiabatic mode is wider than that of the non-adiabatic mode.

An additional feature of the new McStas component is the capability to add a random time to the state of the neutron. 
MIEZE is explicitly a time-dependent method, but most of the sources available in McStas do not produce neutrons in a time window, that is, the initial time for all neutrons is $t_0=0$ rather than $t_{0}\in(0,t_{max}$). 
As a result, the time of arrival at the detector is correlated with the neutron wavelength, as opposed to reality at a constant wavelength source with ``white'' beam being incident on the detector at all times. 
A parameter is included in the RF flipper component which sets $t_{max}$.
A time $t_0$ is added to the McStas time variable for a neutron entering the flipper, where $t_0$ is sampled from a uniform distribution in the range (0, $t_{max}$). 
By setting $t_{max}$ non-zero for the first flipper and zero for all subsequent flippers, this is equivalent to having a random initialization time at the source, removing the correlation between time of arrival at the detector and the wavelength.

\subsection{McStas model of the MIEZE setup}
As shown in eq.~(\ref{eqn:TauMIEZE}), the Fourier time can be varied by changing the sample to detector length $L_s$, the wavelength $\lambda$, or the frequency difference between the RF spin flippers $\Delta\omega=\omega_2-\omega_1$. Changing $L_s$ would also change the variance in the path length from the sample to detector, so it is preferable for this study to set it to a constant. Changing $\lambda$ will also change the scattering vector $\boldsymbol{Q}$ which is measured, and so it is favorable to study the maximum Fourier times for a given scattering angle and wavelength of neutron by scanning the difference of $\Delta\omega$. 
A problem with this in practice is that the cancellation of the velocity dependent phase, eq.~(\ref{eqn:MIEZEcond}), requires that the ratio $\omega_2/\omega_1$ be constant based on the geometry of the beamline. 
Due to the Bloch-Siegert shift \cite{Bloch_1946}, the flipper frequencies cannot be set arbitrarily low and there is also a practical limit to the maximum operation frequency, which together limit the range of $\Delta\omega$ achievable with a set geometry. Effectively these limits set the dynamic range of a MIEZE setup.

\begin{figure}[ht]
	\centering
	\includegraphics[width=0.8\linewidth]{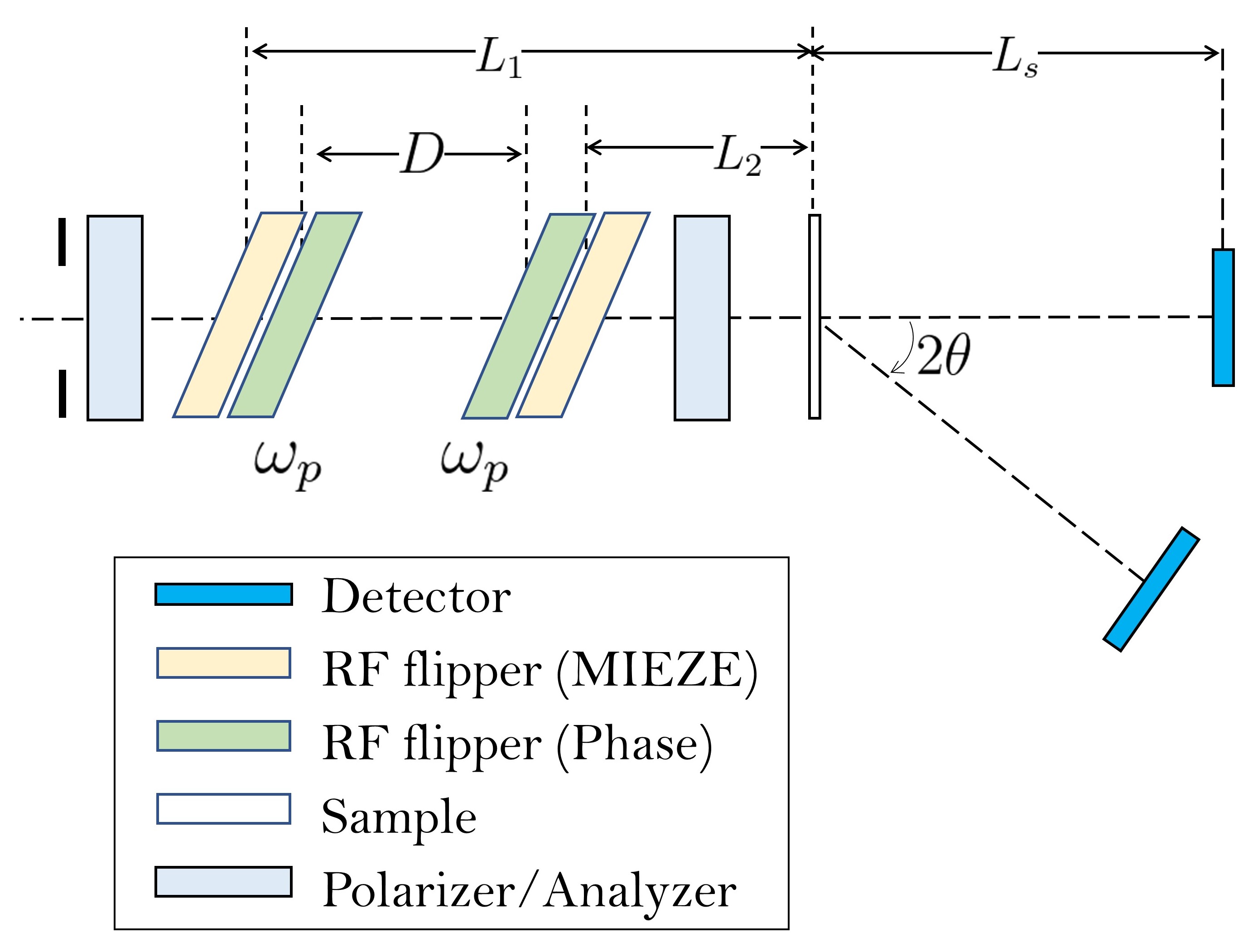}
	\caption{\label{fig:MIEZEsimSchem} Schematic of the components used in the MIEZE simulation, where $L_1=4$~m, $L_2=2$~m, $L_s=2$~m, and $D$ is set so that the phase flippers are spaced at 1~cm inside the MIEZE flippers. The angular frequencies of the phase flippers are equal, $\omega=\omega_p$.}
\end{figure}

It was demonstrated that the velocity dependent phase can also be subtracted by using a guide field similar to NSE in between the RF flippers, increasing the dynamic range by several orders of magnitude at RESEDA~\cite{jochum2019}. 
To explore and maximize the dynamic range of the setup, instead of using a static guide field, we use an additional pair of RF spin flippers operated with the same frequency between the MIEZE RF flippers, which we call phase flippers. 
In this configuration, the frequency difference of the two MIEZE flippers can be set independently without being bound by eq. (\ref{eqn:MIEZEcond}). 
The phase aberration due to the wavelength dispersion can be cancelled by operating the phase flippers at the frequency
\begin{equation}
\omega_p =\frac{1}{D}\cdot\left[ \omega_2 (L_2+L_s)-\omega_1 (L_1+L_s)\right] {\;},
\label{omegap}
\end{equation}
where $D$ is the distance between the phase flippers, and the RF flipper lengths are assumed to be much smaller than $D$. 
With the RF flippers we have tested previously~\cite{Dadisman2020}, this method has since been demonstrated experimentally with detector frequencies in the $10$~mHz-$800$~kHz range for a single geometry configuration.

A full schematic of the MIEZE setup used for the simulations is shown in Fig.~\ref{fig:MIEZEsimSchem}. 
The source was circular in the beam cross-section with a radius which scaled with sample size and maximum divergence settings, focusing neutrons to the sample. The sample is a disk shape of the generic incoherent and elastic scattering type with a thickness in the beam path of 1~mm and variable radius in the beam cross-section. 
The key dimensions of the beamline were chosen to be close to the existing MIEZE beamline RESEDA at the Heinz Maier-Leibnitz Zentrum~\cite{RESEDA}. 
The RF flippers have a dimension similar to  refs.~\cite{LI2020, Dadisman2020} along the beam direction of 10~cm when not tilted. 
The detector has a $10\times{10}$~cm$^2$ active area, and for simplicity we assume an infinitesimally thin detector. 
This is a reasonable simplification for this study, because at large scattering angles the signal at the detector generally has a period greater than 50~$\mu$s. The intensity modulations on the detector is integrated spatially across the detector such that the data can be represented as a 1-dimensional modulation in the time domain. The contrast for each simulation is determined by performing a simple sine fit to the intensity as a function of time for 2 periods of the signal in 20 total time bins. 
For simulations testing the effect of the RF flipper angle, the sample angle is set to $\alpha=0$, and for simulations testing the effect of the sample angle, the RF flipper angle is set to $\beta=0$.

\subsection{Tilting the RF flippers}
\begin{figure}[ht]
\centering
\includegraphics[width=0.6\linewidth]{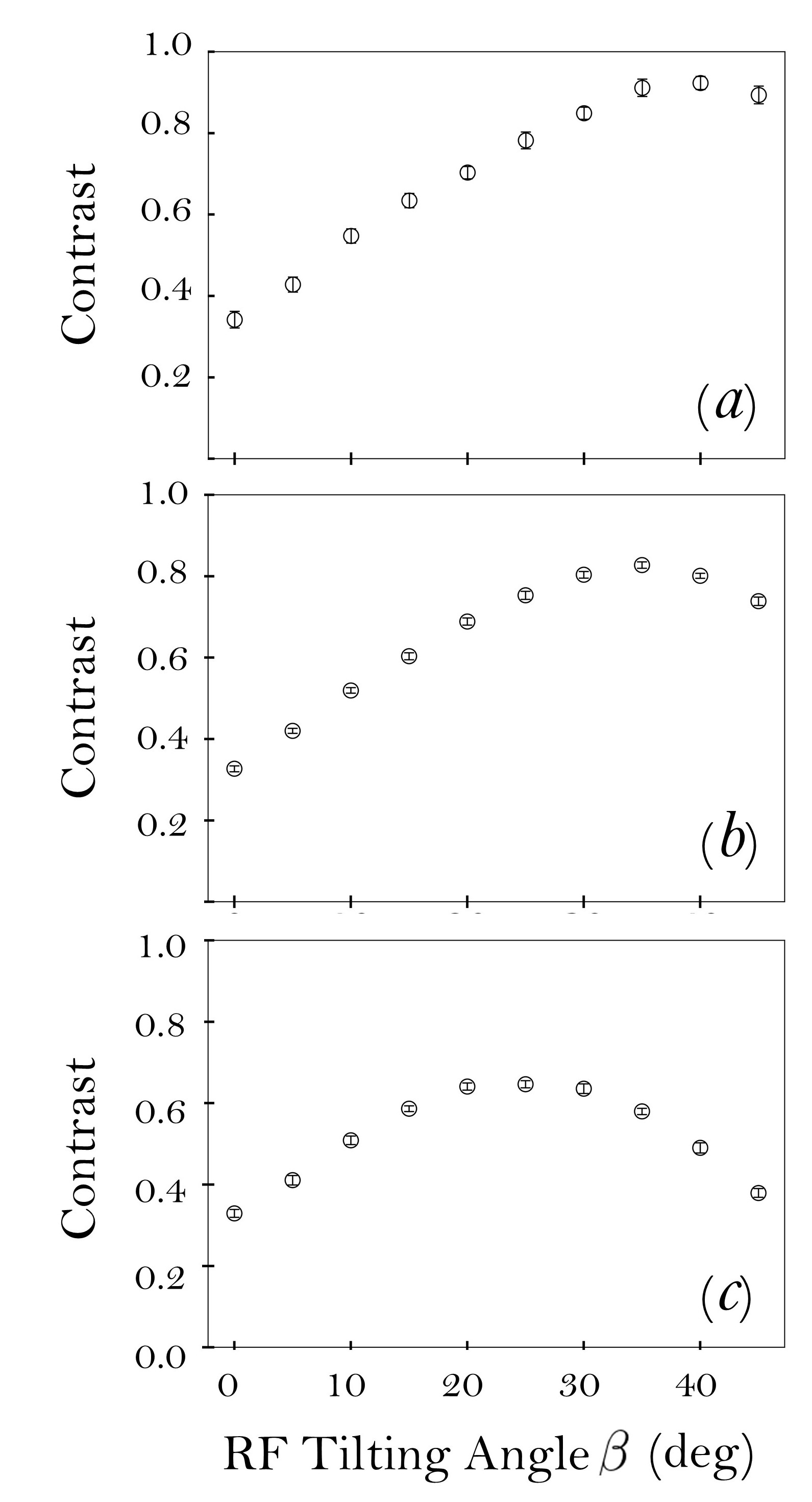}
	\caption{\label{fig:Contrast_v_RFangle} Contrast scan of the intensity modulation at a scattering angle $2\theta=60^{\circ}$ as a function of the RF flipper angle for a maximum divergence of (a) $0.1^{\circ}$, (b) $0.33^{\circ}$, and (c) $0.67^{\circ}$, full width at half maximum (FWHM). The frequency was 10~kHz. From eq.~(\ref{eqn:OptRF}), the maximum contrast is expected at $\beta\approx{41}^{\circ}$.}
\end{figure}

Initial scans of the contrast as a function of the RF flipper angle $\beta$ indicate that tilting the RF flippers does indeed increase the contrast at large scattering angles. 
However, the analytical calculation of eq.~(\ref{eqn:OptRF}) assumed a parallel beam (no divergence) and so it is prudent to study the effect of beam divergence. 
Fig.~\ref{fig:Contrast_v_RFangle} demonstrates that for very low divergence, eq.~(\ref{eqn:OptRF}) is correct, whereas for larger beam divergence the optimum angle $\beta$ becomes smaller and the maximum contrast becomes lower. 
The reason for this effect is that the frequency difference between the two RF flippers introduces modulations in the time domain whereas tilting the RF flippers generates a modulation in the space domain. Because both modulations are directly tied to the RF flippers, their focal planes are coupled with each other. With the MIEZE condition satisfied, it means the focal planes in both the time and space domains superimpose at the detector position. Consequently, as discussed in ref. ~\cite{WP}, the wave front of the time modulation at the sample position for a given scattering angle cannot be precisely shaped, which increases the aberrations at increasing beam divergence ~\cite{WP}. 

\begin{figure}[ht]
\centering
\includegraphics[width=0.8\linewidth]{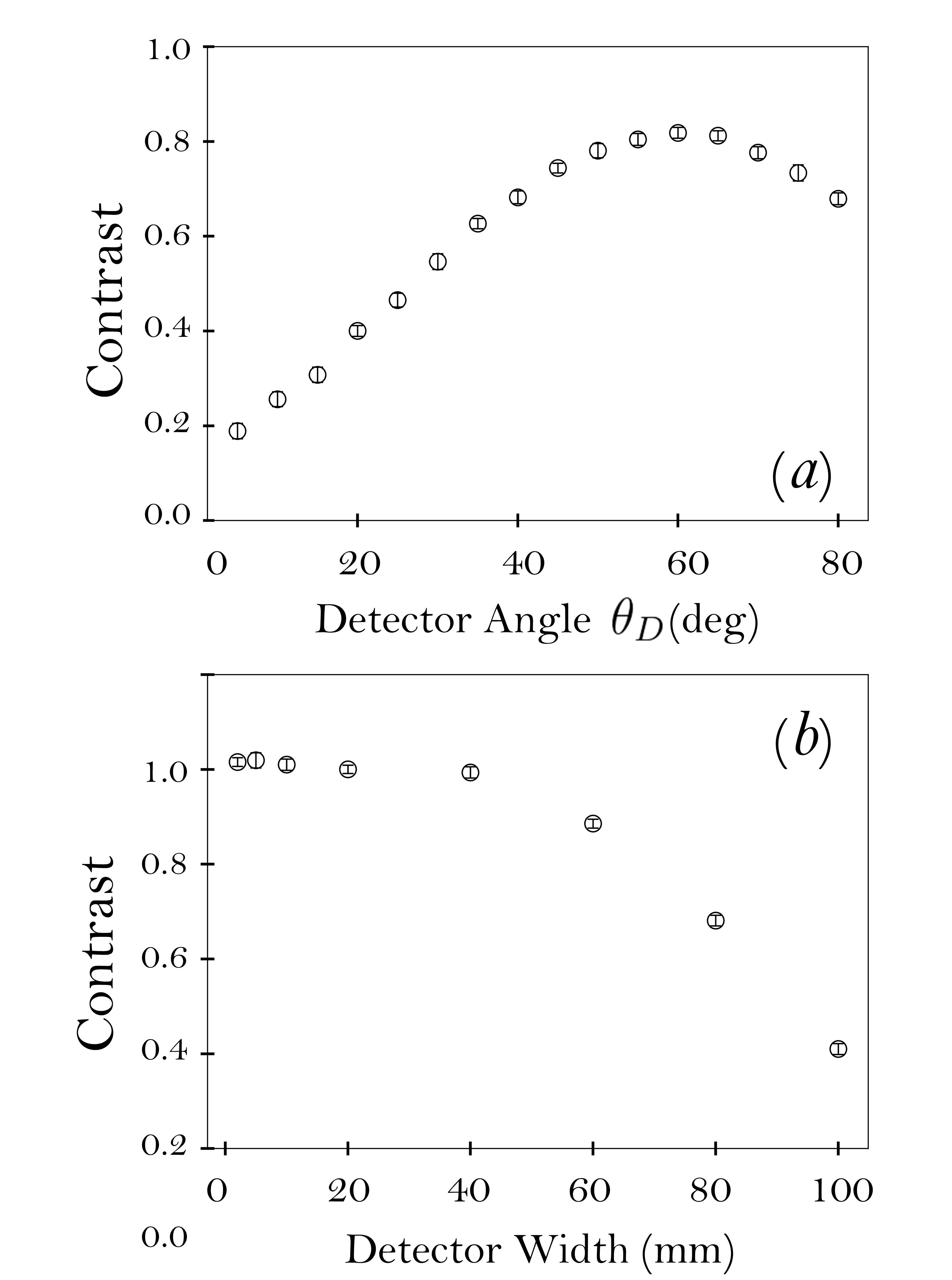}
\caption{\label{fig:ScanDet} Scan of the contrast as a function of (a) detector angle for a width of 10~cm and (b) detector width with $\theta_D=2\theta$=60$^{\circ}$ (FWHM). The maximum incident beam divergence is $0.33^{\circ}$ on a sample with 2~cm radius.}
\end{figure}

While the simulations in Fig.~\ref{fig:Contrast_v_RFangle} used a detector angle of $\theta_D=2\theta$, the work in ref.~\cite{MARTIN2018} found that it is better to maintain $\theta_D=0$ in the SANS regime. Therefore, scans were performed in our setup with tilted RF flippers to see how the detector orientation affected the contrast in a realistic situation. As shown in Fig.~\ref{fig:ScanDet}, it appears that $\theta_D=2\theta$ does maximize the contrast at large scattering angle. Additionally, it is shown that for this configuration the maximum contrast is not strongly dependent on the detector width until the detector is larger than 40~cm. 
The reason for this relative insensitivity to the detector size is that the optimum RF angle only varies by a few degrees across the detector in the scattering plane and the detector takes a small phase volume. When operated at smaller scattering angles and higher frequency, the setup will have a stronger sensitivity to the detector size.

A series of simulations were performed to determine how much the maximum Fourier time could be improved, for a given scattering vector $\boldsymbol{Q}$, by optimally tilting the RF flippers. 
This is done by scanning the difference between the MIEZE RF flipper frequencies, observing how the contrast $P=A/C$ at the detector is reduced with increased Fourier time, and determining the point $\tau_{\mathrm{max}}$ where $P(\tau_{\mathrm{max}})=1/e$. 

\begin{figure}[ht]
\centering
\includegraphics[width=0.65\linewidth, trim={0in 0in 0in 0in}, clip]{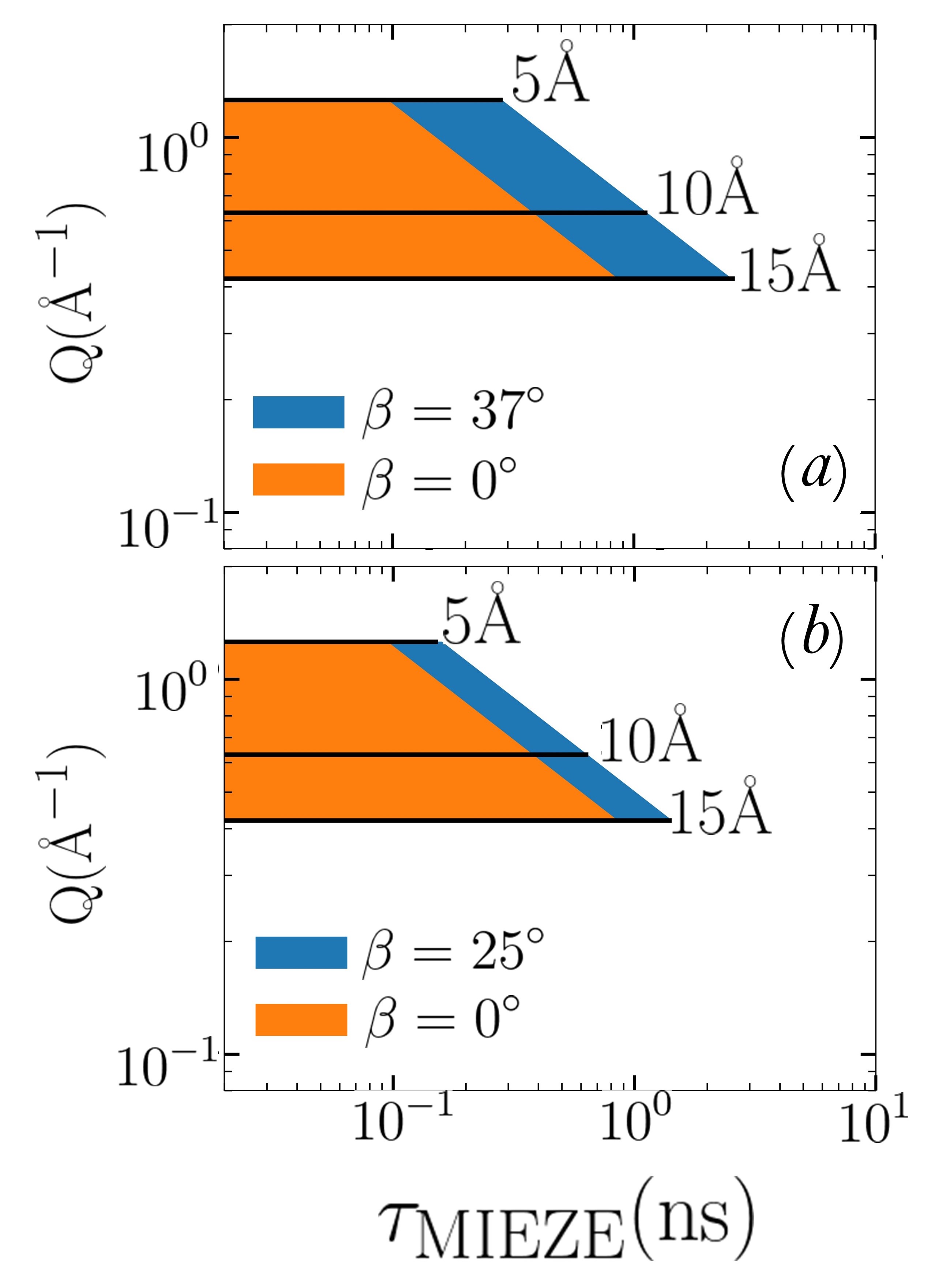}
\caption{\label{fig:Tau_v_Q} Comparison of the maximum Fourier time $\tau_{\mathrm{max}}$ for a given scattering vector $\boldsymbol{Q}$ at a scattering angle of $60^{\circ}$ from a 2~cm radius sample for incident beam divergences of (a) $0.33^{\circ}$ and (b) $0.67^{\circ}$ (FWHM). 
}
\end{figure}

Fig.~\ref{fig:Tau_v_Q} shows the increase in Fourier time that can be achieved with a 2~cm radius sample with two different incident beam divergences. In the figure, the minimum Fourier times are cutoff for visibility, as only the highest Fourier times are of interest here. For a $0.33^{\circ}$ beam divergence, the maximum Fourier time is significantly increased by a factor of 3 by tilting the RF flippers. By comparison, at $0.67^{\circ}$ incident divergence the gain in Fourier time is a little less than a factor 2. Therefore, the balance between neutron flux and beam collimation plays a crucial role in the potential improvement in Fourier time with this method. 
We have also confirmed that there is less to gain when the sample is smaller, but more when the sample is larger, which is expected. 
For example, when the sample radius is reduced to 1~cm in our chosen configuration, the contrast gained by tilting the RF flippers to optimum angles is a factor 1.7 for $0.33^{\circ}$ divergence and 1.2 for $0.67^{\circ}$ divergence.

\subsection{Tilting the sample}
\begin{figure}[ht]
\centering
\includegraphics[width=0.7\linewidth]{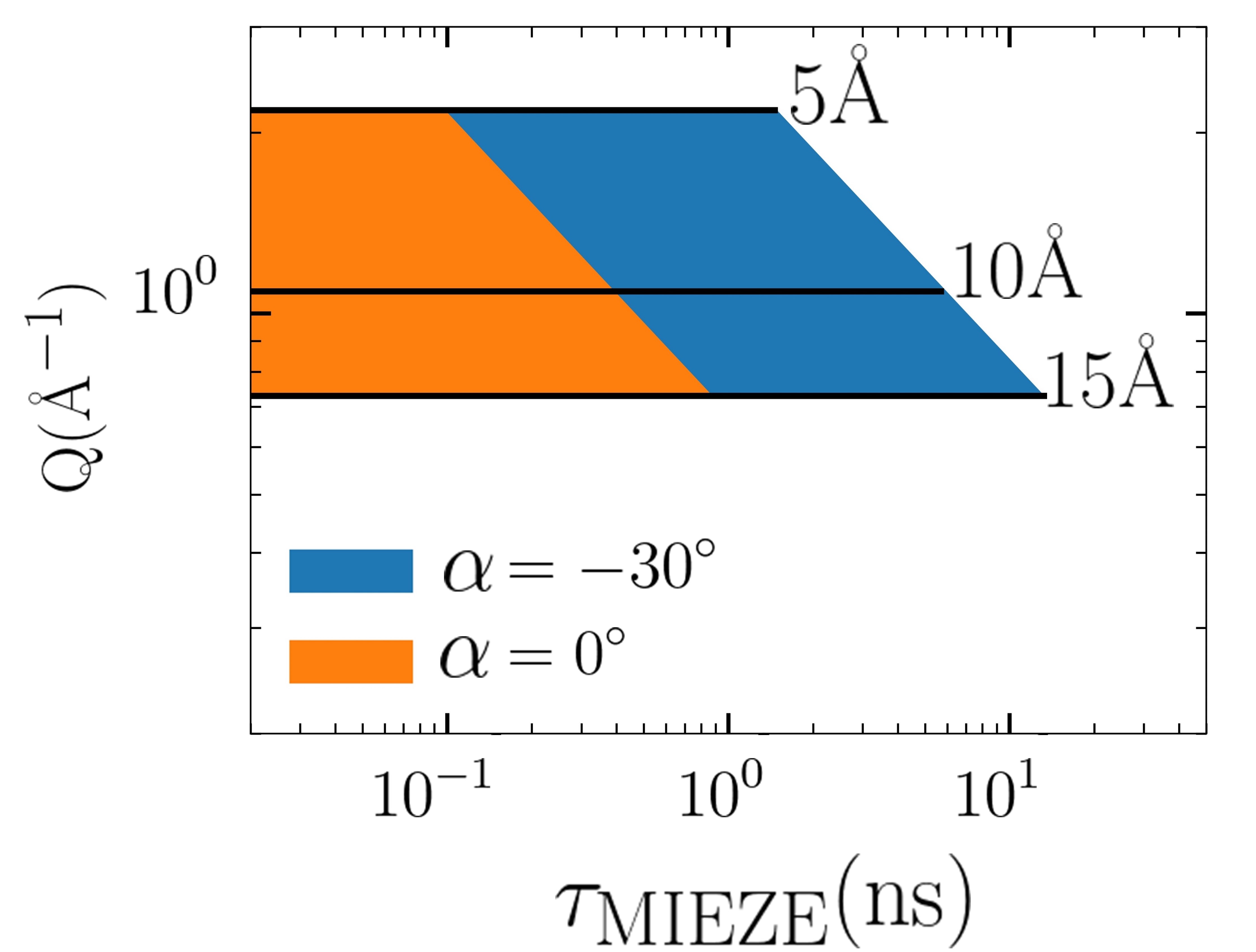}
\caption{\label{fig:SampRot} Comparison of the maximum Fourier time $\tau_{\mathrm{max}}$ for a given scattering vector $\boldsymbol{Q}$ at a scattering angle of $120^{\circ}$ for incident beam with maximum divergence of $0.67^{\circ}$.}
\end{figure}

The increase in achievable Fourier time was also studied for a case where the scattering angle is very large, $2\theta=120^{\circ}$, and the flat disk-shaped sample is rotated to reflection geometry. 
The simulation setup is the same as before, except now the RF spin flippers are not tilted ($\beta=0^{\circ}$), and the sample is tilted in the scattering plane by an angle $\alpha$. Fig.~\ref{fig:SampRot} shows the results for a scattering angle $2\theta=120^{\circ}$ and incident beam with maximum divergence of $0.67^{\circ}$. 
Here, the theoretical best sample angle is $\alpha=-30^{\circ}$. 
Tilting the sample results in an even more significant increase in the maximum Fourier time, a factor of 15 increase for a 2~cm radius sample. We note that this configuration is not strongly affected by the beam divergence, by observing a minimal contrast reduction up to $2^{\circ}$ divergence.

\section{Discussion}

We have shown that it is possible to substantially increase the achievable Fourier time with the MIEZE technique, by tilting the RF spin flippers in the incident beam. 
The proposed modification to an existing experimental setup is minor, and enables one to go away from the SANS regime with MIEZE. 
As such, it promises to be an impactful improvement for MIEZE. 
The effectiveness of the proposed modification increases with larger samples, which provides an additional benefit as it will help to increase the count rate in an experiment. 
However, a large beam divergence will significantly decrease this benefit again. 
We have also demonstrated that, alternatively, at very large scattering angles the sample may be tilted instead of the RF flippers.  
While this was originally considered for the SANS regime~\cite{BRANDL2011,MARTIN2018}, 
we have shown that a substantial increase in the Fourier time may be achieved this way at large angles. 
The best practice may be to tilt the RF flippers for scattering less than $90^{\circ}$ and to tilt the sample for larger scattering angles.

\section{Acknowledgments}
This work is sponsored by the Laboratory Directed Research and Development Program of Oak Ridge National Laboratory, managed by UT-Battelle, LLC, for the U. S. Department of Energy. This material is based upon work supported by the U.S. Department of Energy, Office of Science, Office of Basic Energy Sciences under contract number DE-AC05-00OR22725.

\bibliographystyle{elsarticle-num-names}
\bibliography{LargeAngleMIEZE}

\end{document}